\documentclass[epjST]{svjour}

\usepackage{graphics}
\usepackage{amsmath}
\usepackage{amsfonts}
\usepackage{amssymb}
\usepackage[numbers,sort&compress]{natbib}

\renewcommand{\exp}[1]{\mathrm{e}^{#1}}
\renewcommand{\vec}[1]{\mathbf{#1}}
\newcommand{\unit}[1]{\widehat{\vec{#1}}}

\begin{document}
\title{A model for rolling swarms of locusts}
\author{Chad M. Topaz\inst{1}\fnmsep\thanks{\email{topaz@usc.edu}} \and Andrew J. Bernoff\inst{2}\fnmsep\thanks{\email{ajb@hmc.edu}} \and Sheldon Logan\inst{3}\fnmsep\thanks{Current address: Dept. of Computer Engineering, University of California, Santa Cruz, CA, 95064} and Wyatt Toolson\inst{2}}
\institute{University of Southern California, Los Angeles, CA, 90089, USA
\and Harvey Mudd College Dept. of Mathematics, Claremont, CA, 91711 USA
\and Harvey Mudd College Dept. of Engineering, Claremont, CA, 91711 USA}
\abstract{We construct an individual-based kinematic model of rolling migratory locust swarms. The model incorporates social interactions, gravity, wind, and the effect of the impenetrable boundary formed by the ground. We study the model using numerical simulations and tools from statistical mechanics, namely the notion of H-stability. For a free-space swarm (no wind and gravity), as the number of locusts increases, it approaches a crystalline lattice of fixed density if it is H-stable, and in contrast becomes ever more dense if it is catastrophic. Numerical simulations suggest that whether or not a swarm rolls depends on the statistical mechanical properties of the corresponding free-space swarm. For a swarm that is H-stable in free space, gravity causes the group to land and form a crystalline lattice. Wind, in turn, smears the swarm out along the ground until all individuals are stationary. In contrast, for a swarm that is catastrophic in free space, gravity causes the group to land and form a bubble-like shape. In the presence of wind, the swarm migrates with a rolling motion similar to natural locust swarms. The rolling structure is similar to that observed by biologists, and includes a takeoff zone, a landing zone, and a stationary zone where grounded locusts can rest and feed.}
\maketitle

\section{Introduction}

Biological swarms provide fascinating examples of natural pattern formation on short time scales, and on longer time scales may have significant ecological and environmental consequences \cite{tk1998,ogk2001}. The most dramatic example, arguably, is that of locusts, which cause famines worldwide. Of particular interest are species such as the African migratory locust \emph{Locusta migratoria migratorioides} and the desert locust  \emph{Schistocerca gregaria}, whose habitats together cover the vast majority of northern Africa, the Middle East, and southwestern Asia \cite{u1977}. These locusts, like many others, exhibit an intriguing phase polymorphism. Individuals in the \emph{solitarious} phase avoid social contact. In contrast, adult locusts in the \emph{gregarious} phase form flying swarms. These swarms may contain up to $10^{10}$ members, cover cross sectional areas of up to $1000\ km^2$, and travel up to $10^2\ km$ per day for a period of days or weeks as they feed \cite{u1977} causing devastating crop loss \cite{j1997}. The mechanism for the switch to the dangerous gregarious phase is complex, and has been a subject of significant biological inquiry. A suite of factors recently has been implicated, including fractal geometry of the vegetation landscape \cite{cdsk1998} and mechanosensory stimulus of the locusts' back legs \cite{sdhd2001}. In this paper, we focus on a group of insects already in the gregarious phase and build a mathematical model for the destructive flying swarms.

A migratory swarm develops from gregarious individuals in several stages. First, the insects may form organized bands that march along the ground; these have recently been studied in \cite{bschdms2006}. The marching group become a flying group through a complicated process whose details depend on environmental conditions including time of day, wind, sunshine, and air temperature \cite{k1951,u1977}. In general, locust swarm formation begins with grounded individuals performing short, local flights and other movements which are uncoordinated with those of neighbors. In later stages, groups of locusts take shape, with each group's members sharing a common spatial orientation. Finally, these groups become coordinated with each other, and the swarm development culminates with a mass departure. As the swarm propagates, it forms a rolling motion \cite{a1967,u1977} which is the focus of this paper.

\begin{figure}[t]
\centerline{
\resizebox{\textwidth}{!}{\includegraphics{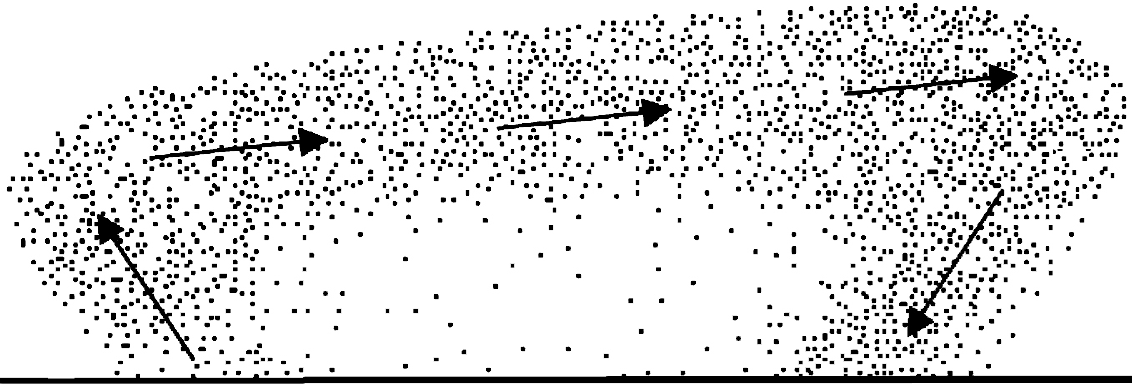}}
}
\caption{Schematic depiction of a rolling locust swarm, after \cite{u1977}. See text for a description. The downwind direction is to the right.}
\label{fig:swarmpic}
\end{figure}

Detailed accounts of the macroscopic swarm motion and the motion of individuals within the group are given in \cite{k1951,a1967,u1977,r1989}. Here we give a brief summary. A schematic diagram is shown in Figure \ref{fig:swarmpic}. The macroscopic direction of the swarm is aligned with the wind. Individual locusts within the swarm move in the following manner. First, flying locusts head downwind towards the front of the swarm. Locusts reaching the front perform a mass landing, heading downwards and slightly upwind until they reach the ground. These locusts remain grounded, resting, feeding, and possibly ovipositing, until the trailing front of the flying swarm passes overhead, at which point they are excited into a massive takeoff upwards and slightly upwind. They eventually turn to align with the wind and again fly towards the front of the swarm until the next landing, and so on.

The only published mathematical models of migrating locust swarms appear in \cite{kwg1998}. This work partitions the locust population into airborne locusts and grounded locusts, and describes each group as a continuum density field depending on a one-dimensional spatial coordinate (aligned with the wind) and on time. The basic model consists of coupled, nonlinear partial differential equations for the two fields. These equations conserve the total number of locusts and account for the continual flux between the grounded and airborne populations. Several variations on the model are considered; these account for random motion and drift of flying locusts, slow motion of locusts on the ground, density dependent turning, and nonlocal social interactions of attractive-repulsive type (see, \emph{e.g.}, \cite{mkbs2003}). The questions of interest is whether there exist traveling band solutions, which consist of a compactly supported region of population density that propagates cohesively over time. The analytical and numerical studies reveal that no such solutions exist, and the authors conclude that their models are insufficient to describe migrating swarms.

Inspired by the work of \cite{kwg1998}, we construct a minimal model for rolling swarms with the goal of reproducing the macroscopic group structure and motion. The key ingredient in our model is the explicit inclusion of vertical structure and the barrier formed by the ground. The added spatial structure enables the model to support cohesive, rolling swarms akin to the ones seen in nature. One of our primary results is a prediction of what parameters allow for this rolling solution, which we accomplish by considering the statistical mechanical properties of the swarm. The swarms found in our model not only roll, but have zones of landing, resting, and takeoff similar to the biologically observed ones.  In addition to presenting a model for rolling swarms, our work addresses the general issue of a swarm interacting with a boundary, which to our knowledge has not been previously considered.

The rest of this paper is organized as follows. In Section \ref{sec:model}, we construct our full mathematical model for rolling locust swarms. The model accounts for social interactions, gravity, and advection in the direction of the wind. Section \ref{sec:freespace} examines a swarm experiencing social interactions in free space, in the absence of gravity and wind. This problem is equivalent to one well-studied in the field of statistical mechanics \cite{r1969}, and has also been examined in the context of biological swarming \cite{mkbs2003,dcbc2006}. We review the relevant results, which depend crucially on whether the parameters of the social interaction potential are chosen to be in the \emph{H-stable} or \emph{catastrophic} regime. In the H-stable regime, individuals form a crystalline group whose size grows as the population increases. The energy per particle of the system approaches a constant which we estimate numerically. In the catastrophic regime, individuals form a closely packed group whose size remains more or less constant. The energy per particle decreases linearly, and we find analytical expressions for bounds on this quantity. In Section \ref{sec:gravity} we incorporate gravity into the model and study it via numerical simulations. When the social interactions are in the H-stable regime, the locusts form an (approximately) crystalline group on the ground. On the other hand, in the catastrophic regime, they form a bubble-like shape with a layer of insects on the ground, a group in the air, and a void space in between.  In Section \ref{sec:wind} we add wind into the model, and see that in the catastrophic case, a rolling swarm forms, similar to those described in \cite{u1977}. In Section \ref{sec:discussion} we discuss our model and our results vis-a-vis biological observations of locusts, and in Section \ref{sec:conclusions} we conclude and mention some directions for future investigation.

\section{Mathematical model}
\label{sec:model}

We classify our model as an individual-based, kinematic, deterministic one. By \emph{individual-based}, we mean that we track the motion of each individual locust \cite{s1973b,ss1973,osik1977,vcbcs1995,gct2001,lrc2001,set2001,ckjrf2002,eea2002,ah2003,ee2003,gct2003,mkbs2003,pvg2003,gc2004,dcbc2006} rather than constructing a continuum density field \cite{k1978,o1980,my1982,dd1984,i1984,a1985,i1985,sm1985,in1987,hm1989,go1994,kwg1998,tt1998,fglo1999,mk1999,sr2002,sr2002b,b2004,tb2004,tsa2004,tbl2006,edll2007,cdmbc2007}. By \emph{kinematic}, we mean that the equation of motion for each locust is first order in time, so that the velocity is simply a function of the locust positions. This approach contrasts with the case where each locust follows Newton's law, so that the equation of motion is second order in time. Stated differently: we neglect inertial forces in our model (see the discussion in \cite{mkbs2003} as well as the models in \cite{vcbcs1995,kwg1998,mk1999,gct2001,ckjrf2002,gct2003,gc2004,tb2004,tbl2006}). The validity of this second assumption will be discussed at greater length in Section \ref{sec:discussion}. Finally, our model is deterministic. Though stochasticity  plays an important role in many swarming systems \cite{ee2003,ee2003b,kes2004,eem2005,see2005}, we neglect the random motion of locusts. We briefly consider noise in Section \ref{sec:discussion}, where we see that it does not qualitatively impact our results.

We now build our model. There are $N$ locusts in the group, and the $i^{th}$ locust has position $\vec{x}_i$. The general model is
\begin{equation}
\label{eq:ge1}
\dot{\vec{x}}_i = \vec{S}_i + \vec{v}_g + \vec{v}_a.
\end{equation}
The (noninertial) forces acting on each locust are the social interactions $\vec{S}_i$, gravity $\vec{v}_g$, and downwind advection $\vec{v}_a$, each of which we discuss below. The direction of swarm migration is strongly correlated with the direction of the wind \cite{u1977,r1989} and has little macroscopic motion in the transverse direction. For simplicity, as in \cite{kwg1998}, we ignore the transverse direction, and so our model is two-dimensional, \emph{i.e.}, $\vec{x}_i = (x_i,z_i)$. Here $x$ is a coordinate which is locally aligned with the main current of the wind and $z$ is the usual vertical coordinate. The explicit inclusion of vertical structure in the model is a critical difference between this work and the work in \cite{kwg1998}.


Two locusts in isolation exert forces on each other according to basic biological principles of attraction and repulsion (see, \emph{e.g.}, \cite{bf1999b,mk1999,ogk2001,mkbs2003}). Repulsion operates very strongly over a short length scale in order to prevent collisions. Attraction is weaker, and operates over a longer length scale, providing the social force necessary for aggregation. For a review of evolutionary aspects of attraction, including increased survival due to the selfish-herd effect, see \cite{ogk2001}. We model the strength of these social forces using the function
\begin{equation}
\label{eq:morse}
s(r) = F \exp{-r/L} - \exp{-r}.
\end{equation}
Here, $r$ is a distance, $F$ describes the strength of attraction, and $L$ is the typical attractive length scale. We have scaled the time and space coordinates so that the repulsive strength and length scale are unity. We assume that $F < 1$ and $L > 1$ so that repulsion is stronger and shorter-scale, and attraction in weaker and longer-scale. This is typical for social organisms \cite{mkbs2003}. The social force exerted by locust $j$ on locust $i$ is
\begin{equation}
\vec{s}_{ij} = s(r_{ij}) \unit{r}_{ij}
\end{equation}
where $r_{ij} = |\vec{x}_j - \vec{x}_i|$ is the distance between the two locusts and $\unit{r}_{ij} = (\vec{x}_j - \vec{x}_i)/r_{ij}$ is the unit vector pointing from $\vec{x}_i$ to $\vec{x}_j$. We take the total social force on each locust in the swarm to be the superposition of all of the pairwise interactions,
\begin{equation}
\label{eq:S}
\vec{S}_i = \sum_{\substack{j=1 \\ j\neq i}}^{N} \vec{s}_{ij}.
\end{equation}
In the field of statistical mechanics, (\ref{eq:morse}) is known as a Morse-type interaction. It has also been used in previous studies of swarming \cite{mkbs2003,dcbc2006}. In the biological setting, each exponential term arises from assuming a constant rate of transmission failure of sensory data per unit distance (the constant hazard function assumption). However, we do not intend for our choice of the Morse interaction to be taken too literally. It is a convenient phenomenological model which, with appropriately chosen parameters, captures the essential features of attraction and repulsion.

\begin{figure}[t]
\centerline{
\resizebox{\textwidth}{!}{\includegraphics{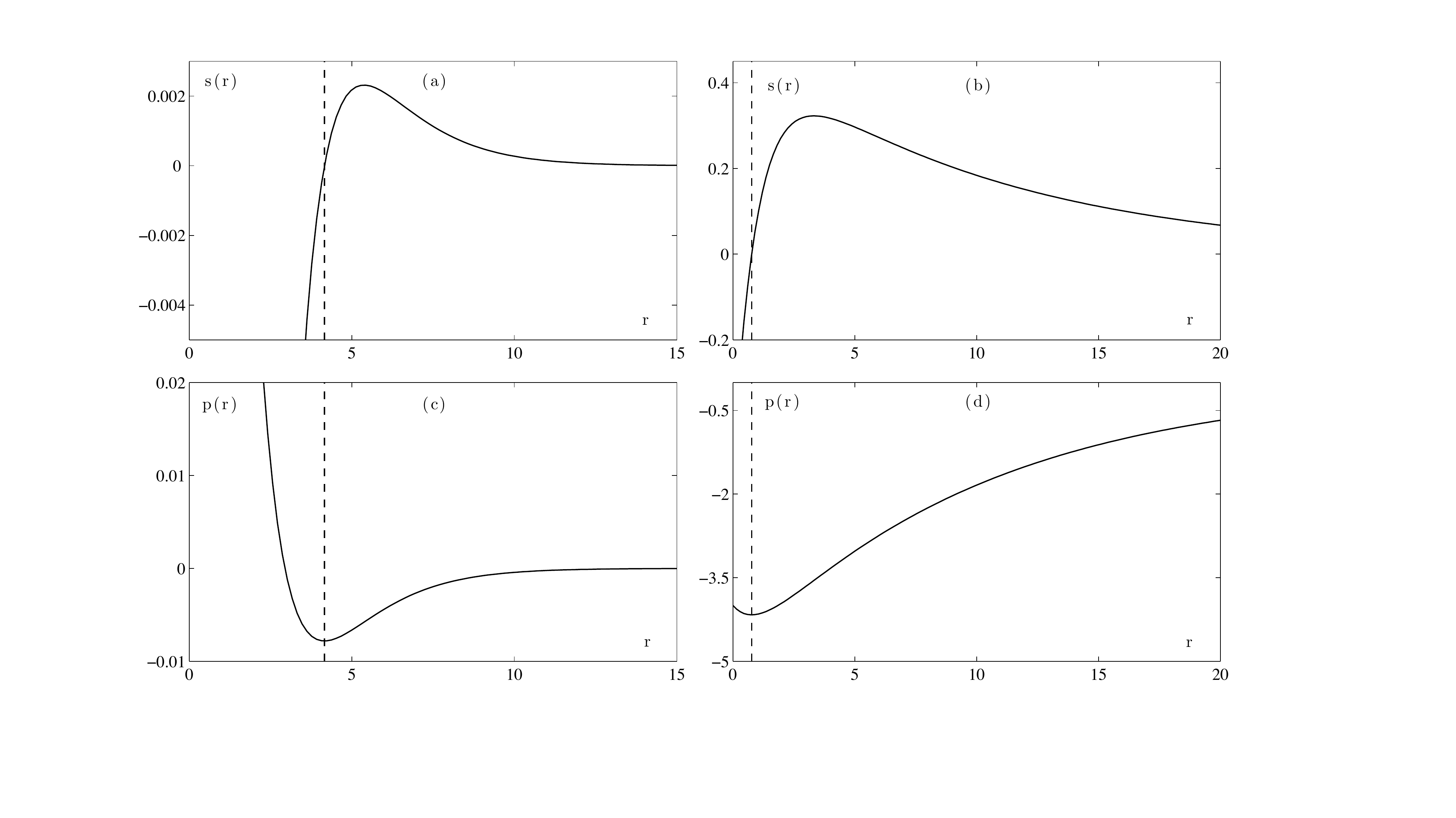}}}
\caption{(a) and (b) Social force $s$ in (\ref{eq:morse}) exerted by one locust on another as a function of their distance r. The broken vertical line indicates the comfortable distance, which is $r_c \approx 4.2$ in (a) and $r_c \approx 0.77$ in (b). (c) and (d) The corresponding potentials $p$ in (\ref{eq:potential}), where the vertical line now indicates the minimum. For (a) and (c), $F=0.25$ and $L=1.5$, which corresponds to the H-stable regime. For (b) and (d), $F=0.5$ and $L=10$, which corresponds to the catastrophic regime. For both cases, $s(r)$ and $p(r)$ are finite as $r \rightarrow 0$.}
\label{fig:potentials}
\end{figure}

See Figure \ref{fig:potentials} for examples of (\ref{eq:morse}) with different parameters. In Figure \ref{fig:potentials}(a), at short distances, $s<0$ and so repulsion dominates. At the vertical dotted line at $r\approx 6.2$, $s=0$ and so attraction and repulsion balance, and the social interaction induces no movement. Following \cite{mkbs2003}, we refer to this distance as the \emph{comfortable distance}. It is given by
\begin{equation}
\label{eq:rc}
r_c = \frac{L \ln F}{1-L}.
\end{equation}
To the right of the dotted line at $r_c$, $s>0$ and so attraction dominates. As $r \rightarrow \infty$, $s \rightarrow 0^+$ since the locusts are too far away to sense each other, and hence, to interact. For this example, $F=0.5$ and $L=10$. Figure \ref{fig:potentials}(b) is similar, but with  $F=0.25$ and $L=1.5$. The comfortable distance is $r_c \approx 4.2$. Even though both examples have short range repulsion and longer-range attraction that both decay to zero over long distances, there is an important qualitative difference betwen the aggregate swarm behavior observed in these two examples. We postpone its discussion until the next section.

Since our model includes vertical structure, we must consider the effect of gravity. Like all forces in our model, gravity operates without inertia, and hence locusts in the absence of other effects (social interactions and downwind advection) experience free-fall at an effective terminal velocity which we refer to as $G$. Thus, 
\begin{equation}
\label{eq:g}
\vec{v}_g = -G \hat{\vec{e}}_z
\end{equation}
where $\hat{\vec{e}}_z$ is the unit vector in the positive $z$ direction.

As mentioned above, locusts fly more or less in the direction of the wind. Whether the flight is active flight in this direction, or passive advection by the wind is irrelevant in our model (see also the discussion on passive versus active flying in \cite{k1951}). We include a constant drift at speed $U$, and thus the advective velocity is
\begin{equation}
\label{eq:u}
\vec{v}_a = U \hat{\vec{e}}_x
\end{equation}
where $\hat{\vec{e}}_x$ is the unit vector in the positive $x$ (downwind) direction.

Substituting (\ref{eq:S}), (\ref{eq:g}) and (\ref{eq:u}) into (\ref{eq:ge1}) yields our governing equation,
\begin{equation}
\label{eq:ge2}
\dot{\vec{x}}_i = \sum_{\substack{j=1 \\ j\neq i}}^{N} s(|\vec{x}_j - \vec{x}_i|) \frac{\vec{x}_j - \vec{x}_i}{|\vec{x}_j - \vec{x}_i|} -G \hat{\vec{e}}_z + U \hat{\vec{e}}_x
\end{equation}
where $s$ is given by (\ref{eq:morse}) and $G$, $U$, $F$ and $L$ are parameters.

We must also define the behavior of locusts on the ground. We assume a flat, impenetrable ground, and thus grounded locusts may not have a vertical velocity which is negative. Furthermore, since locusts rest and feed while grounded, their motion in that state is negligible compared to their motion in the air. The simplest boundary condition satisfying these criteria is the following. For any locust on the ground, if the vertical component of the velocity as computed on the right side of (\ref{eq:ge2}) is nonpositive, then the total velocity of that locust is set to zero. If the vertical velocity is positive, then no adjustment is performed and the locust is allowed to take off. That is,
\begin{equation}
\label{eq:bc}
\vec{\dot{x}}_i\text{ set to } 0 \text{ if } z_i = 0 \text{ and } \dot{z}_i \leq 0.
\end{equation}

The full model is defined by (\ref{eq:ge2}) and (\ref{eq:bc}). In the next section, we ignore gravity and wind ($G = U = 0$) and review what is known about ``free-space'' swarms.

\section{Free-space swarm}
\label{sec:freespace}

We ignore gravity and downwind advection, setting $G=U=0$ in (\ref{eq:ge2}). The simple aggregation model that results is
\begin{equation}
\label{eq:freespace}
\dot{\vec{x}}_i = \sum_{\substack{j=1 \\ j\neq i}}^{N} s(|\vec{x}_j - \vec{x}_i|) \frac{\vec{x}_j - \vec{x}_i}{|\vec{x}_j - \vec{x}_i|}
\end{equation}
which describes locusts experiencing attractive and repulsive interactions in free space, that is, without other effects and without boundaries. It will be convenient to define a potential $p(r)$ which is an antiderivative of (\ref{eq:morse}),
\begin{equation}
\label{eq:potential}
p(r) = -FL\exp{-r/L}+\exp{-r}
\end{equation}
and note that the system has a free-space energy 
\begin{equation}
\label{eq:Efs}
E_{fs} = \frac{1}{2}\sum_{i=1}^N \sum_{j=1}^N p(r_{ij}).
\end{equation}
Equation (\ref{eq:freespace}) may then be written in gradient form,
\begin{equation}
\label{eq:gradient}
\dot{\vec{x}}_i = - \nabla_i E_{fs}.
\end{equation}
A few lines of straightforward calculation verify that (\ref{eq:gradient}) is equivalent to (\ref{eq:freespace}), and that $E_{fs}$ is an energy ($dE_{fs}/dt \leq 0$).

Equation (\ref{eq:freespace}) is studied in \cite{mkbs2003}, which considers the parameter space of (\ref{eq:morse}) and asks whether individuals form a cohesive group and, if so, what the typical distance between organisms is. Certain conditions on $F$ and $L$ must be satisfied in order to have a cohesive group; in particular, $L$ must be sufficiently large and $F$ must be sufficiently small, so that repulsion dominates at short distances and attraction dominates at large distances. For the elementary case of two organisms, if a cohesive state exists, the distance is just the comfortable distance. For the case of more organisms, the so-called \emph{individual distance} is shown in \cite{mkbs2003} to be shorter than the comfortable distance.

The system (\ref{eq:freespace}) is well-known in the field of statistical mechanics \cite{r1969} and a recent paper \cite{dcbc2006} draws connections to biological swarming. The key notion is one called H-stability which relates to the behavior of the group as a function of the population size $N$, and which depends on the shape of the interaction function $s(r)$. For our choice of $s(r)$ in (\ref{eq:morse}), this reduces to conditions on the parameters $F$ and $L$. We state the conditions that $F$ and $L$ must satisfy for H-stability at the end of this section.

\begin{figure}[t]
\centerline{\resizebox{\textwidth}{!}{\includegraphics{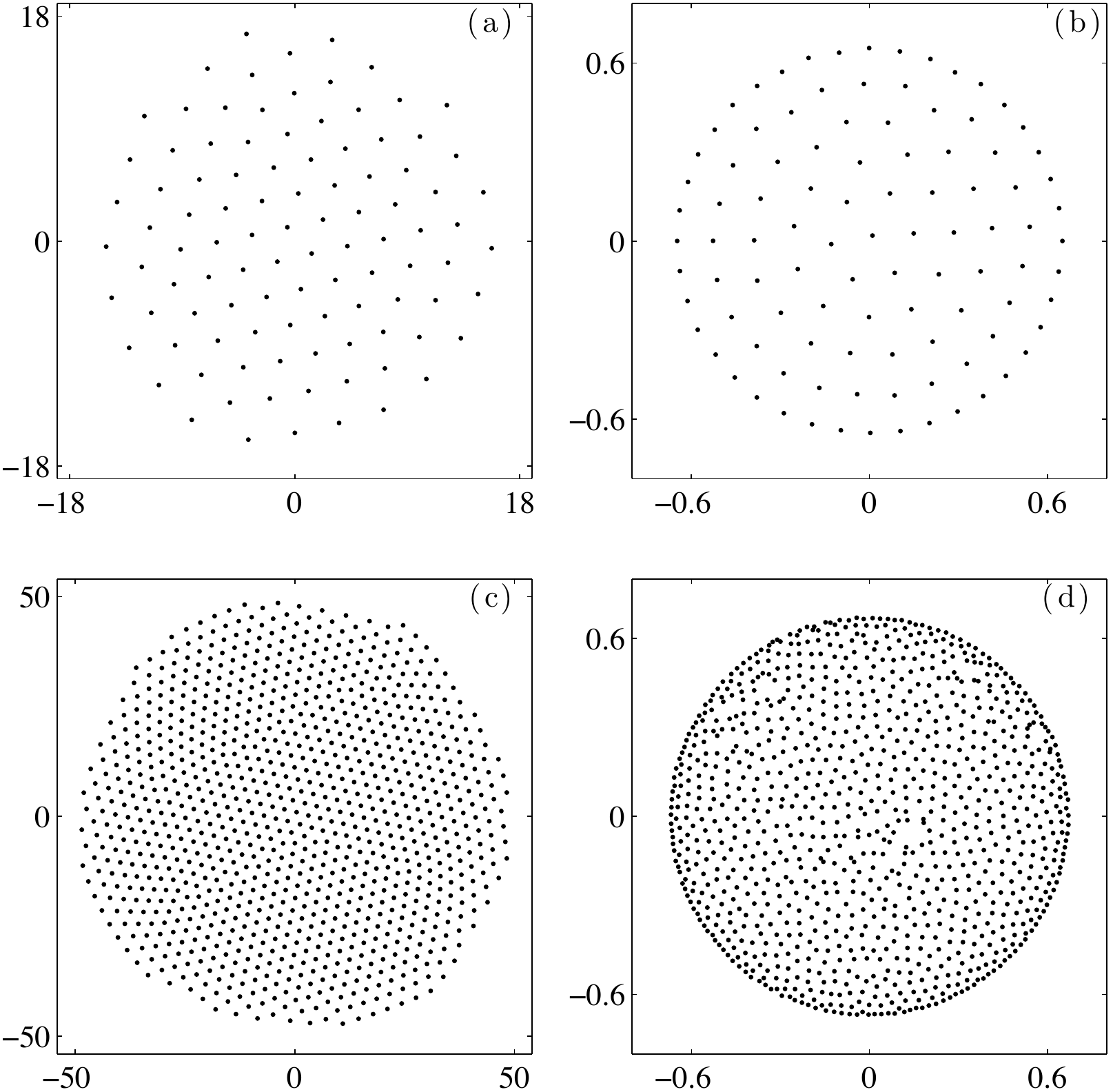}}}
\caption{Equilibrium free-space swarms obtained by numerically minimizing the energy $E_{fs}$ given by (\ref{eq:Efs}). In (a) and (c), $N=100$ and $N=1000$ respectively. The social interaction rule is the one depicted in Figure \ref{fig:potentials}(ac), which corresponds to the H-stable regime. As $N$ is increased, the individual distance is preserved, so the more populous group covers a larger spatial area. In (b) and (d), the same values of $N$ are used, but the interaction rule is the one in Figure \ref{fig:potentials}(bd), which corresponds to the catastrophic regime. In this case, as $N$ is increased, organisms pack together more densely.}
\label{fig:freespace}
\end{figure}

We seek equilibrium solutions by performing an numerical unconstrained nonlinear minimization of the energy $E_{fs}$ in (\ref{eq:Efs}). Given an initial condition, the algorithm finds a single local minimizer of $E_{fs}$ However, as one might expect, multiple minimizers actually exist. We have verified their existence by fixing the model parameters $N$, $F$, and $L$, and using a variety of random initial conditions. The final states achieved are not identical. However, we find that for the values of $N$ we study, the different minimizers are statistically indistinguishable in that their total size and energy are essentially identical. The local minima seem to be tightly clustered in the energy landscape $E_{fs}$.

\begin{figure}[t]
\centerline{
\resizebox{\textwidth}{!}{\includegraphics{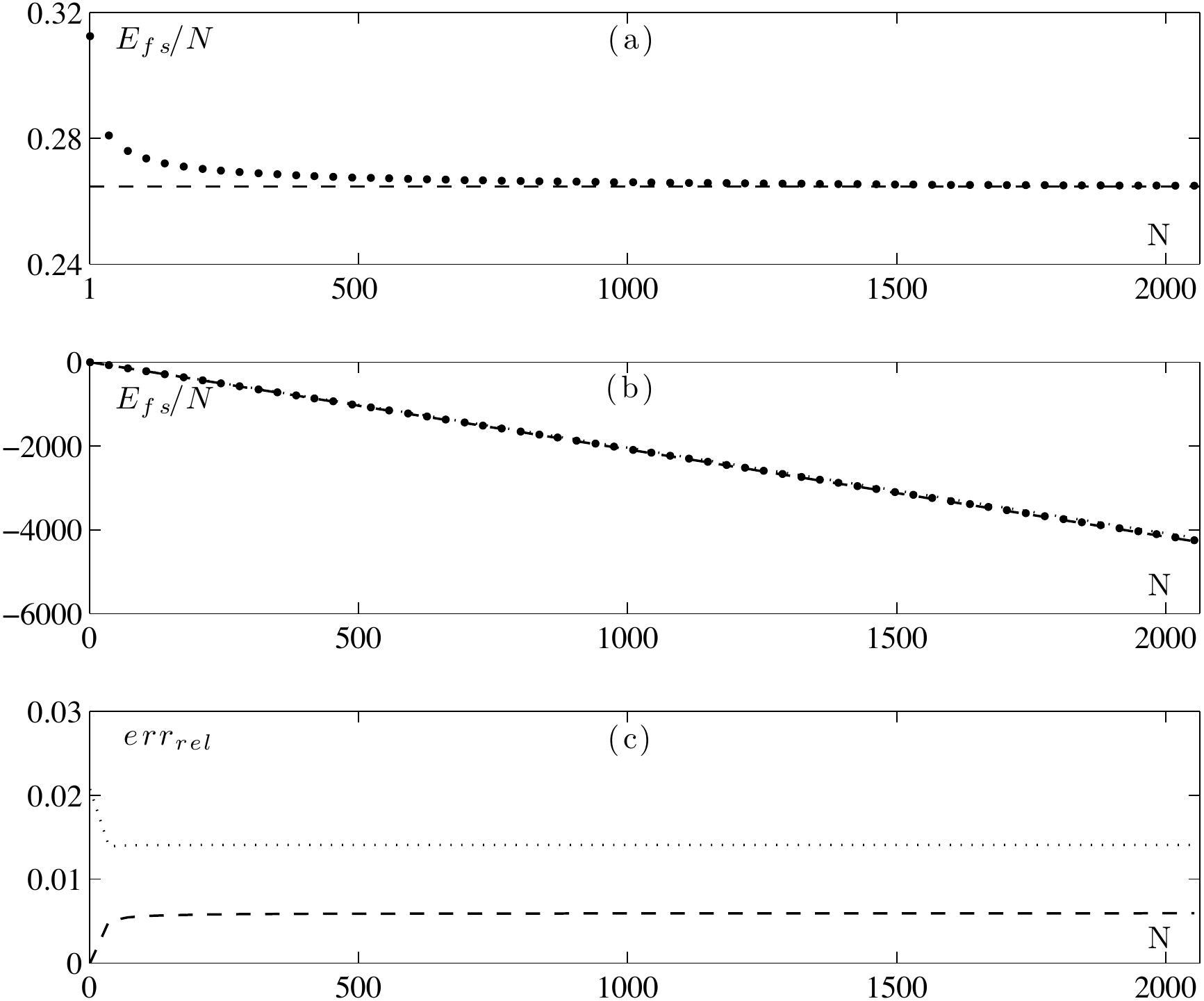}}}
\caption{(a) Energy per locust $E_{fs}/N$ for equilibrium free-space swarms ($G=U=0$ in Eq. \ref{eq:ge2}) in the H-stable regime. The energy is given by (\ref{eq:Efs}), and the social interaction parameters $F,L$ are those from Figure \ref{fig:potentials}(ac). As $N$ increases, the equilibrium configurations have a value of $E_{fs}/N$ that approaches that of a perfect, infinite hexagonal lattice as indicated by the broken asymptote. (b) Like (a), but the parameters are those from Figure \ref{fig:potentials}(bd) which are in the catastrophic regime. $E_{fs}/N$ decreases linearly. Lower and upper bounds are given by (\ref{eq:lb}) and (\ref{eq:ub}). These are shown (respectively) as dotted and broken lines, but are difficult to distinguish from the actual data on this plot. (c) Relative error in using the lower (dotted) and upper (broken) bounds as estimates for the data in (b).}
\label{fig:energyperparticle}
\end{figure}

We now discuss the two important statistical mechanical regimes, following \cite{dcbc2006}. If the parameters $F$ and $L$ are chosen in the \emph{H-stable} regime, then as $N$ increases, the spacing between individuals approaches a finite constant. In the parlance of \cite{mkbs2003}, the group is \emph{well-spaced}. Figure \ref{fig:freespace}(ac) shows equilibria for two swarms with $N=100$ and $N=1000$. The interaction parameters correspond to those in Figure \ref{fig:potentials}(ac). We see that organisms form a crystalline hexagonal lattice where the nearest-neighbor distance is approximately equal for all individuals (excluding edge effects). As more individuals are added to the group, the inter-organism spacing is preserved and the group grows to cover a larger spatial region. The energy per organism $E_{fs}/N$ approaches a constant \cite{r1969} as shown in Figure \ref{fig:energyperparticle}(a). This constant is the value of $E_{fs}/N$ for a perfect, infinite hexagonal lattice, which we have calculated numerically and indicated by the horizontal asymptote.

On the other hand, if the parameters are chosen outside of the H-stable regime, the system is \emph{catastrophic}. In this case, the energy per organism is unbounded as $N \rightarrow \infty$, and organisms pack together more and more closely as the population size increases. The individual distance is not preserved as a function of $N$. Figure \ref{fig:freespace}(bd) shows example equilibria, again with $N=100$ and $N=1000$; the interaction parameters correspond to Figure \ref{fig:potentials}(bd). The area covered by the two swarms is nearly equal. As $N$ increases, so does the density of the group. $E_{fs}/N$ decreases linearly without bound, as shown by Figure \ref{fig:energyperparticle}(b).

To estimate $E_{fs}/N$ for the catastrophic case, we construct bounds as follows. Of the $N^2$ terms inside the summation in  (\ref{eq:Efs}), there are $N$ self-interaction terms equal to $p(0) = 1-FL$. To find a lower bound, we may assume that the remaining $N^2 - N$ terms take on the smallest possible value, namely $p(r_c)$, the value of the potential at the comfortable distance. Even though this physical configuration of locusts cannot be realized, it gives a lower bound on the energy per particle, namely the line
\begin{equation}
\label{eq:lb}
E_{fs}/N \geq\frac{1}{2} \Bigl[ p(0) + (N-1)p(r_c)\bigr] = \frac{1}{2}\Bigl[(1- FL) - F^{\frac{L}{L-1}}(L-1)N\Bigr].
\end{equation}

Since the equilibria minimize the energy at least locally, and since the local minima are tightly clustered in energy space, almost every physically realizable configuration of locusts will serve as an upper bound. We construct the bound by considering a convenient state with $N/2$ locusts superposed at one point and $N/2$ locusts superposed at the comfortable distance $r_c$. There are $N$ self-interactions each contributing $p(0)$. There are are additional $N(N/2-1)$ contributions of $p(0)$ due to interactions of locusts with others at the same site. Finally, there are $N^2/2$ contributions of $p(r_c)$. Thus, an upper bound on the energy per particle is
\begin{equation}
E_{fs}/N \leq \frac{1}{2} \Bigl[ p(0) + \Bigl(\frac{1}{2}N - 1\Bigr)p(0) + \frac{1}{2}Np(r_c)
\Bigr] =  \frac{1}{4}N \Bigl[(1- FL) - F^{\frac{L}{L-1}}(L-1) \Bigr].
\label{eq:ub}
\end{equation}
The lower and upper bounds are shown (respectively) as dotted and broken lines in Figure \ref{fig:energyperparticle}(b). Because the bounds are difficult to distinguish from the actual values of $E_{fs}/N$ on the plot, we have plotted the relative error of the bounds in Figure \ref{fig:energyperparticle}(c). The upper bound calculation is aided by the fact that the total energy of two locusts at a distance $d$ is proportional to $p(0) + p(d)$ which is negative. Our sample state is constructed by gradually superposing locusts on the two original ones, and this leads to the line in (\ref{eq:ub}). If our original ''base'' state were to have positive energy, our upper bound would have been a line with positive slope, which is indeed an upper bound, but is not useful. In this case, one would wish to construct the bound differently. This will be a subject of future work.

The biological relevance of H-stability versus catastrophe is that many organisms are thought to have a preferred spacing (see the review in \cite{mkbs2003}) more or less independent of population size, and this corresponds to H-stability. Some other organisms are conjectured \cite{dcbc2006} to pack more tightly with increasing population size, which would be catastrophic behavior. The implication of H-stability versus catastrophe is one of the main points ultimately explored in this paper.

The conditions on $F,L$ in (\ref{eq:morse}) required for a cohesive and H-stable group are reviewed in \cite{r1969}. We omit the statistical mechanical details here. The condition is
\begin{equation}
\label{eq:hstable}
1 < L <F^{-1/3}.
\end{equation}
If $L > F^{-1/3} > 1$, then a cohesive but catastrophic group will form (if $L < 1$ then repulsion dominates at large distances and a cohesive group will not form). Note that the condition for H-stability is quite restrictive. For instance, consider an attractive length scale $L$ five times as long as the repulsive one. Then for the system to be H-stable, the typical force $F$ due to attraction must be less than $0.8\%$ of the typical repulsive force. We stress that one must consider the statistical mechanical condition (\ref{eq:hstable}) to correctly understand the group. The examples on the left and the right in Figure \ref{fig:potentials} look qualitatively similar, with short range repulsion, long range attraction, and a comfortable distance in between. Nonetheless, the examples in Figures \ref{fig:freespace} and \ref{fig:energyperparticle} show that in the limit of large $N$, the systems have quite different behavior.

\section{Swarm under the influence of gravity and a lower boundary}
\label{sec:gravity}

\begin{figure}[t]
\centerline{
\resizebox{\textwidth}{!}{\includegraphics{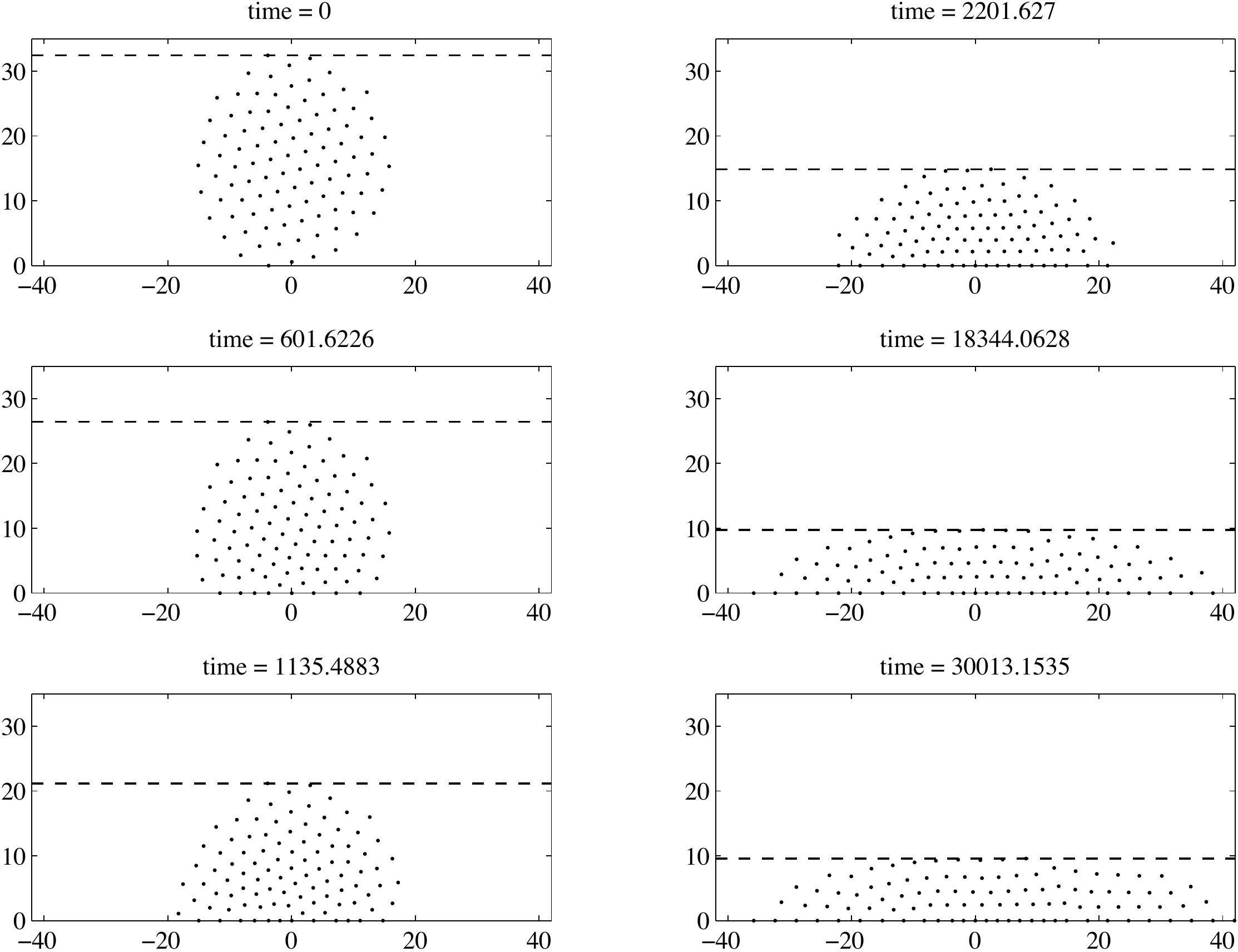}}}
\caption{Snapshots of the H-stable swarm in Figure \ref{fig:freespace}(a) landing on the ground under the influence of gravity. The broken line guides the eye to the top of the swarm. As the swarm lands, the macroscopic behavior is to (approximately) maintain a crystalline structure. Here, $N=100$, the social interaction rule is the one in Figure \ref{fig:potentials}(ac), and the gravity parameter is $G=0.01$.}
\label{fig:hstable_100_landing}
\end{figure}

\begin{figure}[t]
\centerline{
\resizebox{\textwidth}{!}{\includegraphics{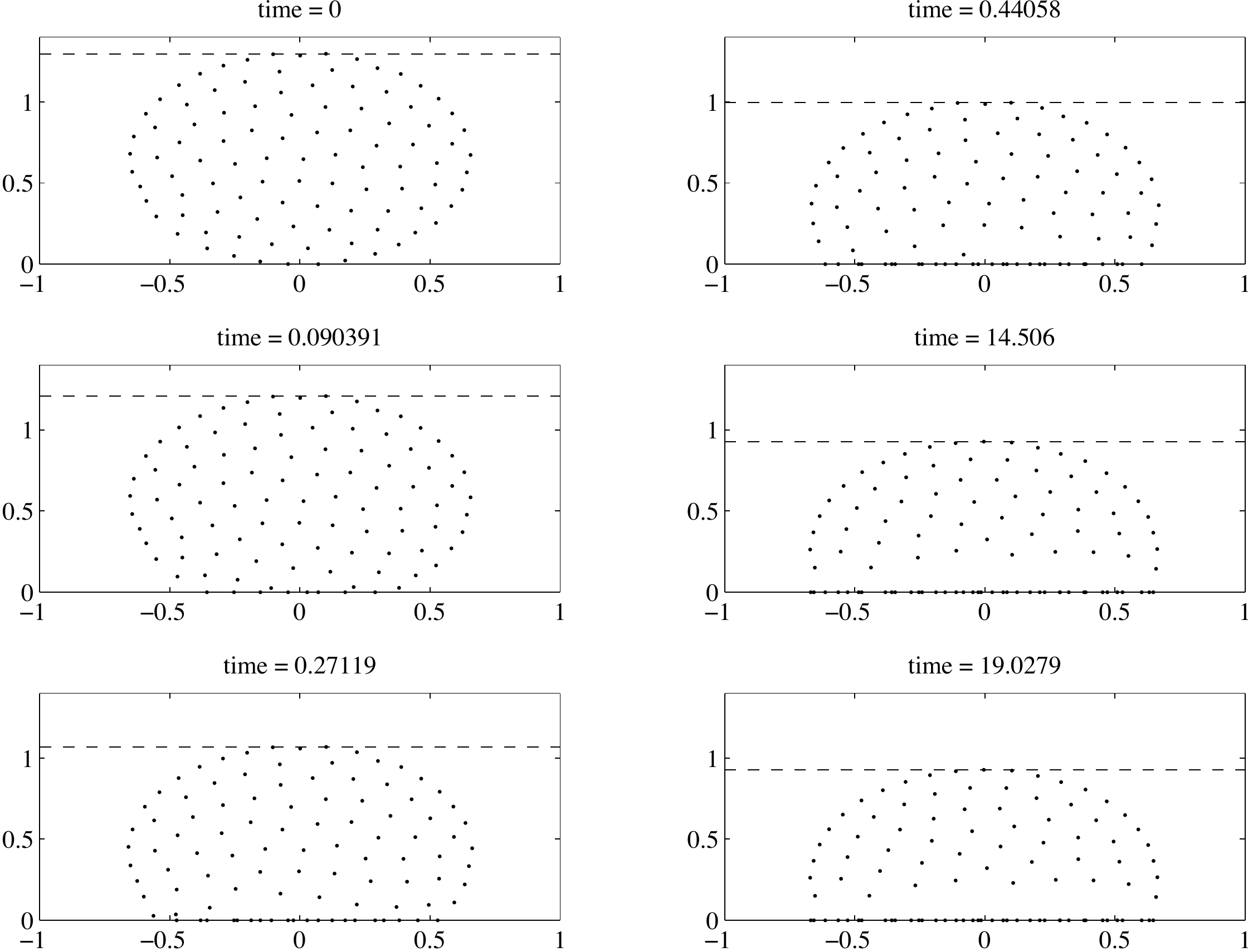}}}
\caption{Snapshots of the catastrophic swarm in Figure \ref{fig:freespace}(b) landing on the ground under the influence of gravity. The broken line guides the eye to the top of the swarm. The equilibrium state consists of a group of locusts in the air, a dense group of locusts on the ground, and a void separating them. Here, $N=100$, the social interaction rule is the one in Figure \ref{fig:potentials}(bd), and the gravity parameter is $G=1$.}
\label{fig:catastrophic_100_landing}
\end{figure}

We expand the investigation in the previous section by considering a swarm in the presence of gravity ($G \neq 0$). If there were no boundary, the system would retain its gradient character, with a modified energy
\begin{equation}
E_g = E_{fs} + \sum_{i=1}^N Gz_i 
\end{equation}
and the system could be converted into the free space system by moving into a
reference frame descending at a speed $G$. However, if we incorporate a horizontal boundary, the swarm descends and some individuals land, as shown in Figures \ref{fig:hstable_100_landing} and \ref{fig:catastrophic_100_landing}. The individuals in free space still have velocities given by $\dot{\vec{x}}_i = - \nabla_i E_g$ but the individuals on the ground are no longer mobile. Note that as an alternative, we could formally include the constraint that $z_i \ge 0$ for grounded locusts by making the energy $E_g$ infinite for $z_i<0$. This approach would allow locusts to  move along the ground with the horizontal component of their free space velocity, and the dynamics would locally minimize the constrained energy $E_g$.

However, as discussed in Section \ref{sec:model}, we choose the more realistic boundary condition (\ref{eq:bc}) since the motion of grounded locusts is minimal. Because this  ``sticky'' boundary condition prohibits horizontal motion of grounded locusts, the problem loses its gradient character. Though the boundary condition destroys the gradient character of the free-space problem, it is still true that $d E_g/dt \leq 0$. The system will evolve \emph{towards} minimizing $E_g$ for the locusts in the air. The locusts on the ground would have a nonpositive velocity if the ground were penetrable. These dynamics are said to be \emph{quasi-gradient} in character.

As the system is quasi-gradient , we expect the final equilibrium state to be highly dependent upon the initial condition.  A simple thought experiment clarifies this. Consider the case where all the locusts are initially randomly distributed on the ground, and consider applying an infinitesimal  perturbation to one locust (only perturbations with positive $z$ component are allowed under Eq. \ref{eq:bc}). The equilibrium is neutrally stable since the locust will merely fall back to the ground, perhaps slightly displaced in the horizontal. Stated differently, all distributions of locusts initially on the ground are equilibrium configurations.

For the examples below, we consider a swarm that is transiently in free-fall with some time to equilibrate towards its free space configuration before individuals land. When we incorporate the effect of wind in the next section, the geometry of the rolling swarm state we observe appears to be insensitive to the choice of initial condition.

\begin{figure}[t]
\centerline{
\resizebox{\textwidth}{!}{\includegraphics{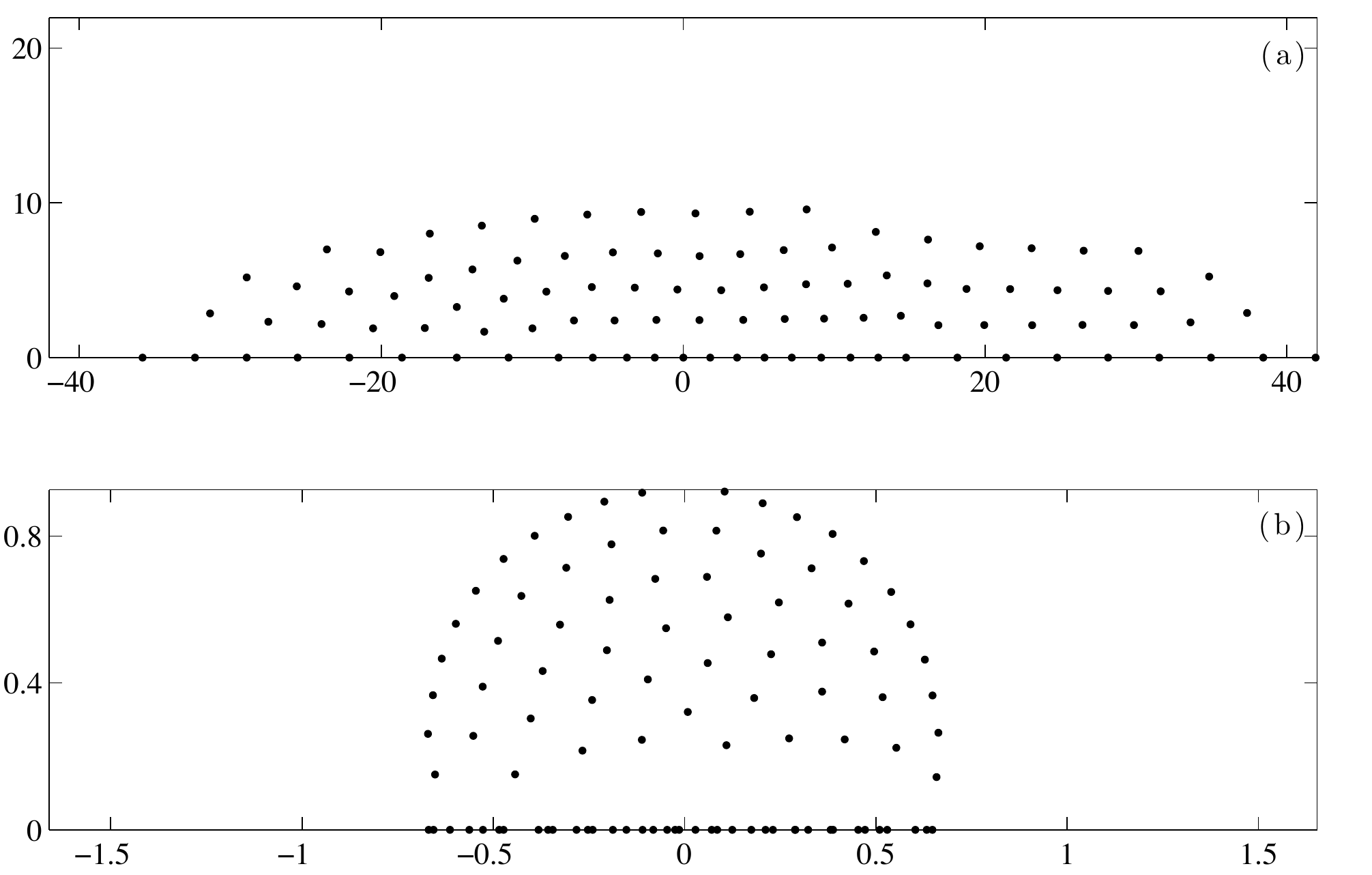}}}
\caption{Blow-up showing final states from (a) Figure \ref{fig:hstable_100_landing} and (b) Figure \ref{fig:catastrophic_100_landing}. Within each sub-plot, the scales on the x and y axes are equal. The locusts land on the ground due to gravity, but the qualitative character of the final state depends on the statistical mechanical regime in which the parameters $F,L$ in (\ref{eq:morse}) lie. In the H-stable case (a) the group has an oblong crystalline structure, while in the catastrophic case (b) it has a bubble-like structure with a void above the ground.}
\label{fig:landed_states}
\end{figure}

Figure \ref{fig:hstable_100_landing} shows the swarm in Figure \ref{fig:freespace}(a) landing under the influence of gravity of strength $G=0.01$. The (initially) nearly circular swarm flattens into an oblong equilibrium configuration on the ground. Due to the H-stable character of the problem, the swarm retains an essentially crystalline structure. In contrast, Figure \ref{fig:catastrophic_100_landing} shows the swarm in Figure \ref{fig:freespace}(b) landing with $G=1$. As the group lands, a dense layer of locusts accumulate on the ground. This dense layer induces a force which is strongly repulsive at short ranges, which in turn leads to a void in the region immediately above the ground. Airborne locusts occupy a region above the void. Overall, the shape of the swarm is bubble-like. Figure \ref{fig:landed_states} compares the final states of the two examples, from which the qualitative difference is apparent.

Numerical investigations for a selected set of values of $F$, $L$ and $G$ other than those used in Figures \ref{fig:hstable_100_landing} and \ref{fig:catastrophic_100_landing} suggest that bubbles only form in the catastrophic case, and never in the H-stable case. The question of exactly what conditions enable a bubble to form is a complex one. It is explored in forthcoming work \cite{bt2007} in which we formulate a continuum description of the swarm and study its energy minimization analytically.

\section{Swarms under the influence of gravity, a lower boundary, and wind}
\label{sec:wind}

\begin{figure}[t]
\centerline{
\resizebox{\textwidth}{!}{\includegraphics{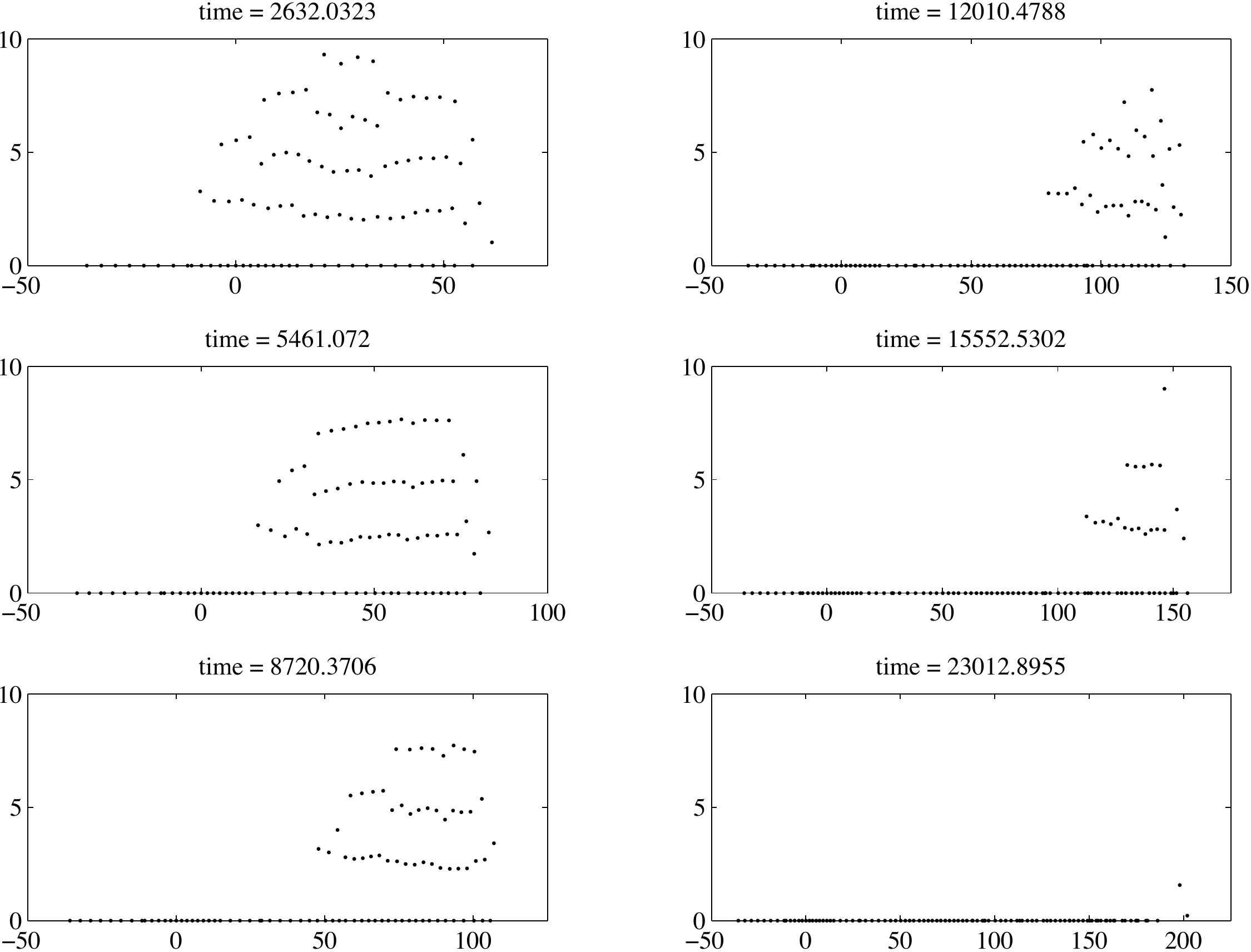}}}
\caption{Evolution of the H-stable state in Figure \ref{fig:landed_states}(a) with wind $U=0.01$. Locusts fly downwind, and those a the front of the group land on the ground, never to take off again. The simulation reaches an equilibrium once the entire group has been smeared  out along the ground.}
\label{fig:smeared_hstable}
\end{figure}

Building on the previous two sections, we now consider the full model (\ref{eq:ge2}) and (\ref{eq:bc}) which incorporates social interactions, gravity, and wind. Figure \ref{fig:smeared_hstable} shows the evolution of the H-stable state in Figure \ref{fig:landed_states}(a) with wind $U = 0.01$. As the group flies downwind to the right, locusts towards the front of the swarm land on the ground, and no locusts on the ground ever take off. After sufficient time, all locusts have landed on the ground and no further movement takes place.

\begin{figure}[t]
\centerline{
\resizebox{\textwidth}{!}{\includegraphics{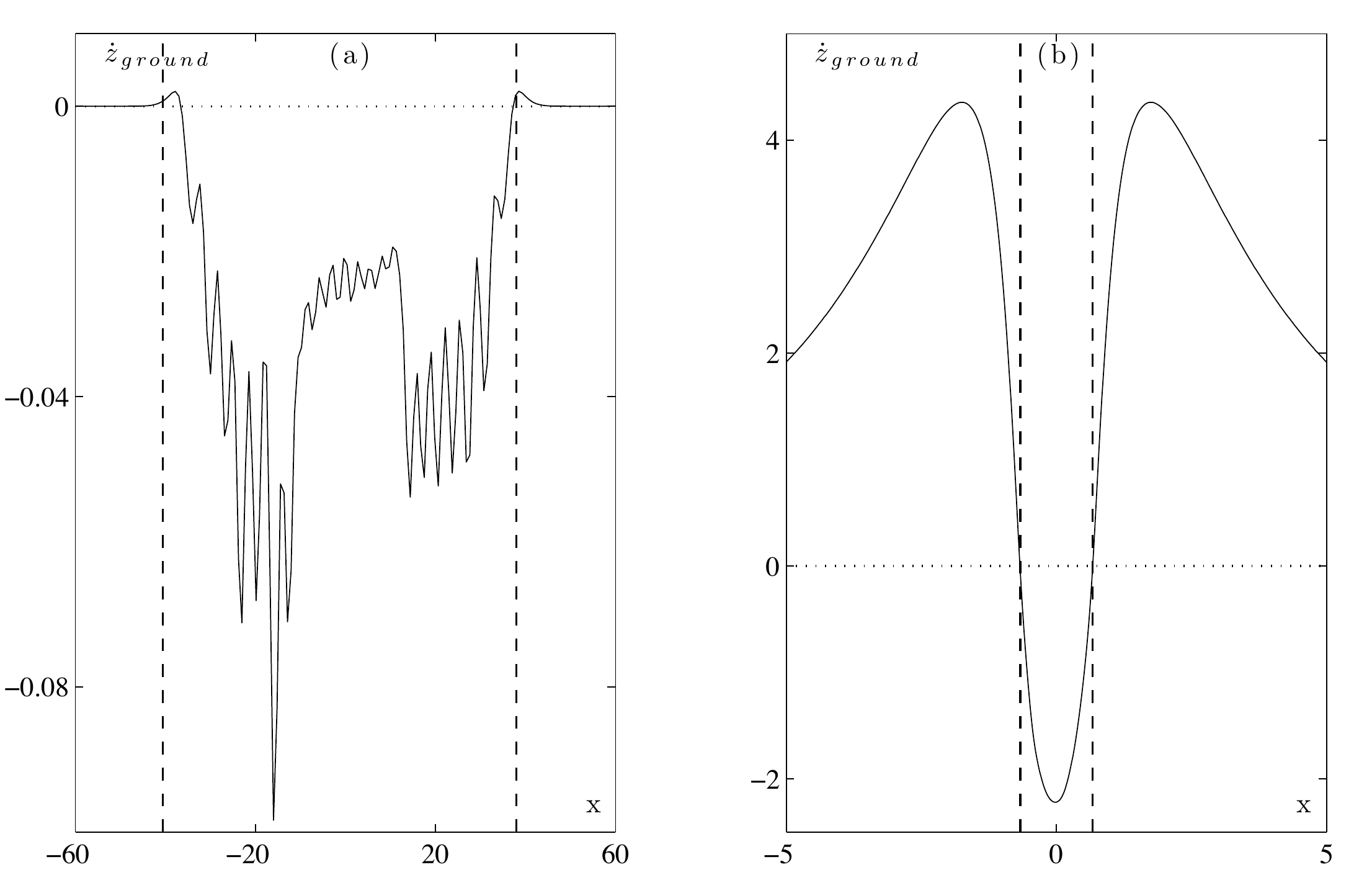}}}
\caption{Vertical component of velocity $\dot{z}_{ground}$ induced by a swarm on a``virtual locust'' positioned on the ground at location $x$. Where $\dot{z}_{ground} < 0$, virtual locusts cannot take flight for any values of the gravity parameter $g$. Where $\dot{z}_{ground}>0$, virtual locusts can take off if $g < \dot{z}_{ground}$. The dotted horizontal line guides the eye to $\dot{z}_{ground}=0$. (a) H-stable case corresponding to the swarm in Figure \ref{fig:landed_states}(a). Virtual locusts outside the swarm boundaries can take off only for miniscule $g$. The broken vertical lines indicate the left and right boundaries of the swarm. (b) Catastrophic case corresponding to the swarm in Figure \ref{fig:landed_states}(b). The swarm induces a large upward velocity on virtual locusts outside of the swarm, suggesting why locusts are not left behind when the swarm migrates.}
\label{fig:ground_velocity}
\end{figure}

This behavior may be understood by considering the social force exerted by the swarm on an imaginary locust, as shown in Figure \ref{fig:ground_velocity}(a). Here we consider the initial state in Figure \ref{fig:landed_states}(a) used for the simulation in Figure \ref{fig:smeared_hstable}. The horizontal axis corresponds to the horizontal position $x$, and the vertical dashed lines indicate the (initial) support of the swarm. The vertical axis is $\dot{z}_{ground}(x)$, the vertical velocity that the swarm induces on a virtual locust placed on the ground at horizontal location $x$.  The dotted line represents the zero threshhold. In most of the support of the swarm, the induced vertical velocity is negative.  Near the edges inside the swarm, there is a small region of very weak positive vertical velocity. Outside the support of the swarm, the vertical velocity is negligible. Thus, a grounded locust that happens to be left behind by even a small amount will be grounded permanently, even for very small values of $G$. Stated differently, the attractive force exerted by the swarm is too weak to induce stragglers to take off.

\begin{figure}[t]
\centerline{
\resizebox{\textwidth}{!}{\includegraphics{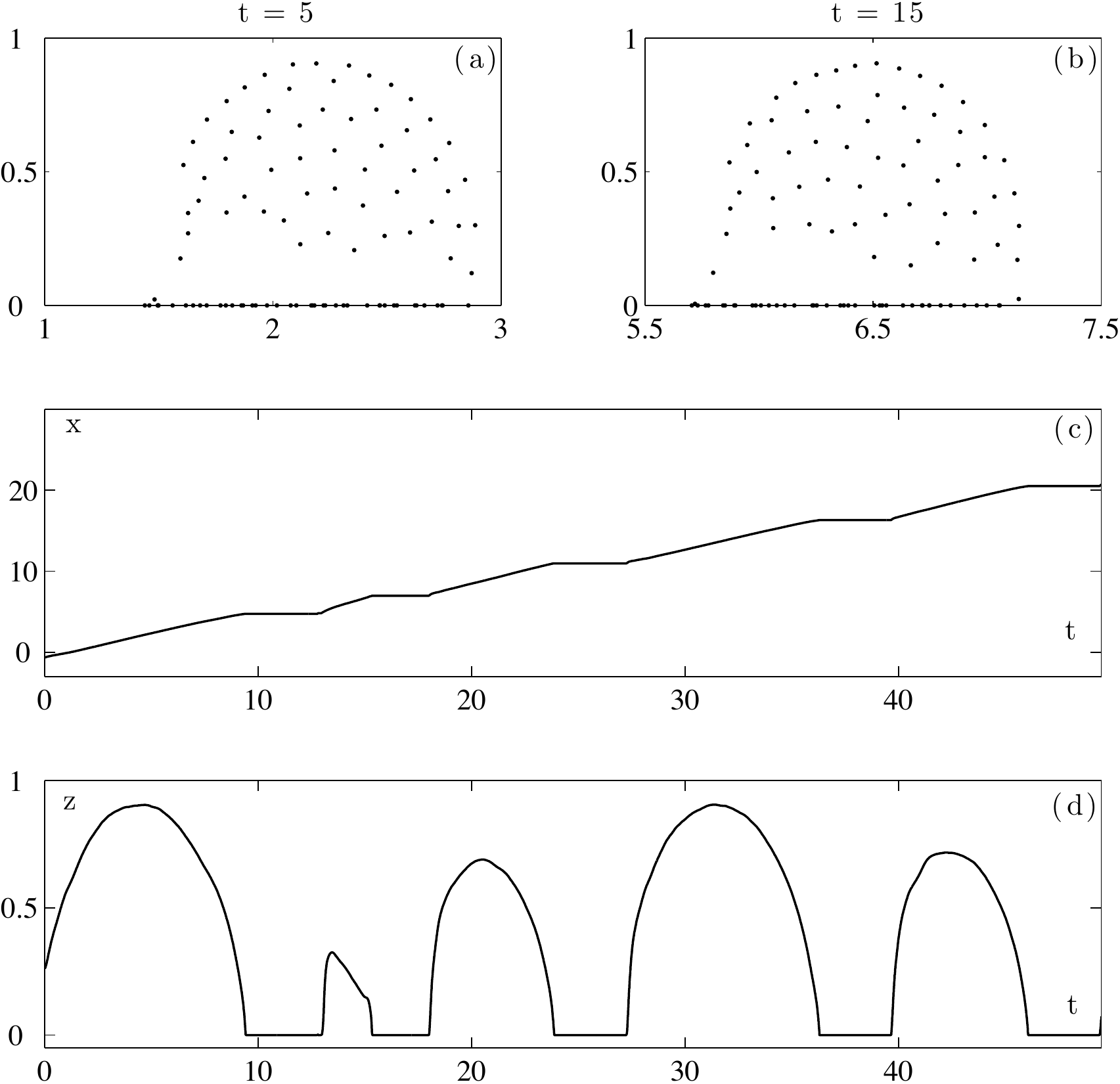}}}
\caption{Evolution of the catastrophic state in Figure \ref{fig:landed_states}(b) with wind $U=1$. The group rolls downwind with locusts at the front of the airborne swarm landing on the ground and those at the back of the grounded group taking flight, as shown in snapshots at times (a) $t=5$ and (b) $t = 15$. Note the qualitative similarity to Figure \ref{fig:swarmpic}. The $x$ and $z$ coordinates of a single locust are tracked in (c) and (d) respectively. When the locust is grounded ($z=0$) there is no horizontal motion due to the boundary condition (\ref{eq:bc}). When the locust is flying, $x$ increases approximately linearly.}
\label{fig:rolling_swarm}
\end{figure}

Figure \ref{fig:ground_velocity}(b) is analogous, but corresponds to the  initial state in Figure \ref{fig:landed_states}(b). In this case, the induced vertical velocity outside of the swarm is strong and positive. A single locust left behind feels a strong upward attractive force, and will take off to rejoin the swarm. Thus, we expect from this plot that the initial state may give rise to a cohesive, traveling swarm if gravity is not too strong. Simulations show that this is indeed the case. With wind $U=1$, the bubble-like state in Figure \ref{fig:landed_states}(b) undergoes a rolling motion to the right. Locusts at the front of the airborne swarm land on the ground, while those at the back of the grounded group take flight. Two snapshots of the rolling motion are shown in Figure \ref{fig:rolling_swarm}(ab); note the qualitative similarity to Figure \ref{fig:swarmpic}. To further understand this motion, we examine time series of $x_i$ and $z_i$ for a single locust, shown in Figure  \ref{fig:rolling_swarm}(cd). The locust flies upwards and downwind, arcing over the void section in the middle of the swarm, with $x_i$ increasing (approximately) linearly in time. The locust reaches a maximum height and then begins descending, with $x_i$ still increasing linearly. The locust settles on the ground in the landing zone at the front of the swarm, and at this point its  horizontal motion stops due to the boundary condition (\ref{eq:bc}); this is seen as the horizontal plateaux in Figure \ref{fig:rolling_swarm}(c). After a time in the stationary zone on the ground, the void and the swarm have passed overhead, and the locust is in the takeoff zone. Nearly left behind by the flying swarm, it leaves the ground in order to catch up.

\begin{figure}[t]
\centerline{\resizebox{\textwidth}{!}{\includegraphics{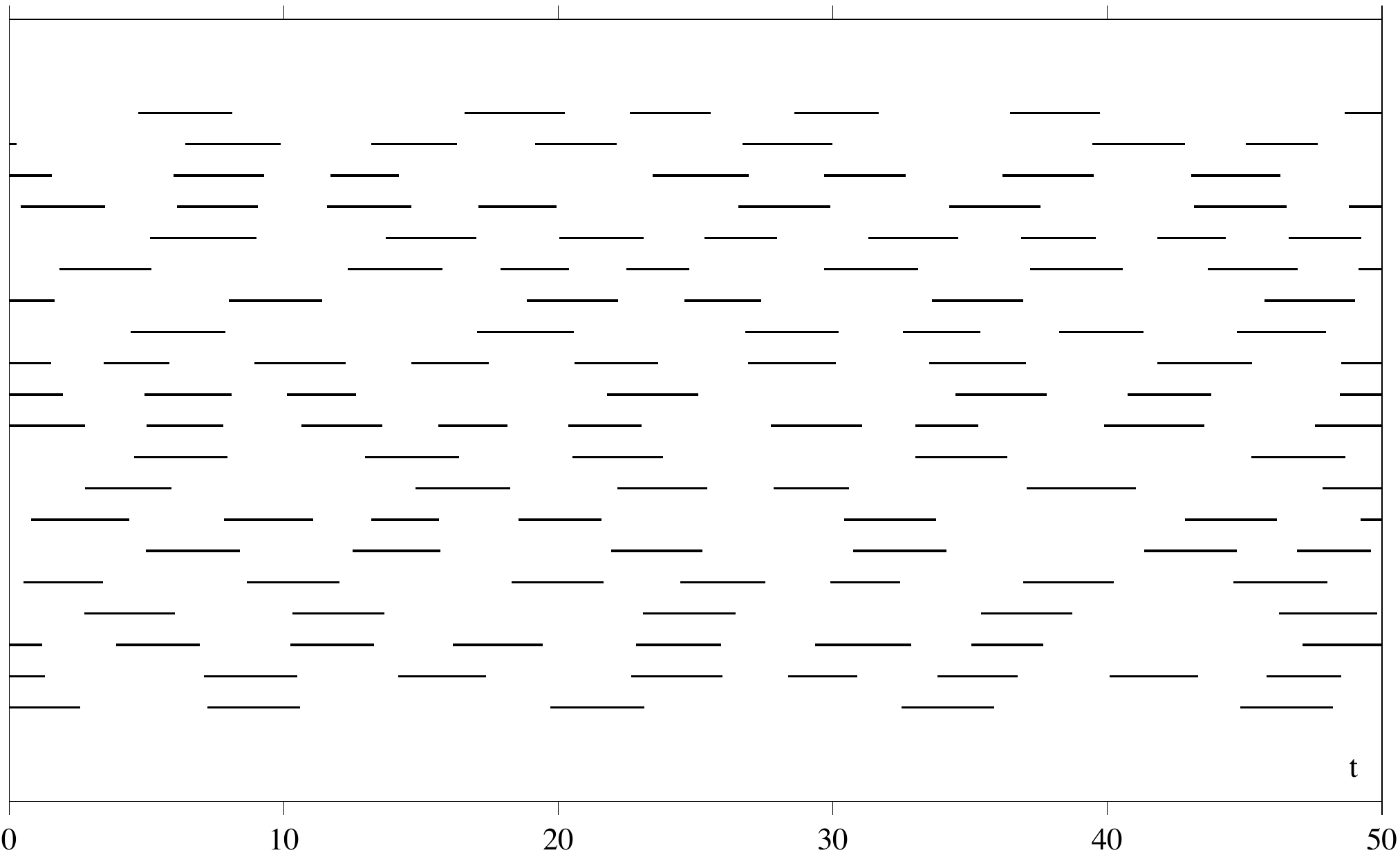}}}
\caption{Visualization of the landing and takeoff pattern of the rolling locust swarm from Figure \ref{fig:rolling_swarm}. The vertical axis indexes $20$ locusts randomly selected from the population. The horizontal axis is time. A darkened pixel represents a time at which a given locust was on the ground. Landed segments with less than 0.3 time units of flight between them are concatenated to be a single segment.}\label{fig:onground}
\end{figure}

One question of interest is whether all locusts undergo motion similar to that described above and shown in Figure \ref{fig:rolling_swarm}(cd). To answer this question, we have computed basic statistics describing the landing/takeoff pattern of the swarm members. This pattern is visualized in Figure \ref{fig:onground}. The plot depicts $20$ locusts randomly selected from the population. The vertical axis indexes the locusts, and the horizontal axis is time. A darkened pixel represents a time at which a given locust was on the ground. For the entire hundred-locust swarm, individuals averaged over the course of the simulation 6.7 segments on ground with standard deviation of 1.1. The mean duration of each grounded segment was 3.0 time units with standard deviation of 0.23. These figures indicate that all locusts follow a similar landing/takeoff pattern. Furthermore, across all times, the average fraction of the population grounded at a given time was 0.39 with a standard deviation of 0.0088. Across all individuals, the average fraction of time spent on the ground was 0.39 (since averaging over locusts commutes with averaging over time) with a standard deviation was 0.049. These statistics confirm that in terms of distribution between the ground and the air, the swarm looks the same at all times.

\section{Discussion}
\label{sec:discussion}

Several aspects of our model and our results bear further discussion vis-a-vis biological observation. First, as discussed in Section \ref{sec:model}, we have used the framework of a kinematic (as opposed to dynamic) model, and thus we have neglected inertial forces acting on locusts. While this may be a good approximation during the bulk of the flying stage, it is likely not accurate during takeoff and landing. Relatedly, takeoff and landing in our simulation occur in the upwind direction, as seen in Figure \ref{fig:rolling_swarm}. However, in the schematic depiction in Figure \ref{fig:swarmpic}, locusts take off and land in the downwind direction. This phenomenon is described in more depth in \cite{u1977}, which cites a combination of behavioral, physiological and aerodynamic factors as explanation. It is possible that incorporating these factors within a dynamic model formulation would lead to simulated swarms with the correct landing and takeoff direction.

Second, the social interactions in our model are of isotropic form, \emph{i.e.}, a given locust senses equally well in every direction, and the influence of any other locust only depends on the distance (and not the angle) between them. Sensing within a swarm is thought to be both visual and auditory \cite{u1977}. Isotropy is a likely a good model for hearing. For sight, we think of it as being a first approximation. Anisotropic interactions have been modeled in a one-dimensional setting in \cite{edll2007}. The effect of anisotropic interactions on locust swarming may be considered in future work.

Third, our model does not incorporate random motion of locusts. In reality, flying locusts will experience random motion due to turbulent air currents and due to the inherent imprecision of their sensing and movement. As a preliminary test of the effect of randomness, we have conducted simulations that include noise added to the right hand side of (\ref{eq:ge2}). For the $i^{th}$ locust, we chose the noise to have direction uniformly distributed from $[0,2\pi]$ and the magnitude uniformly distributed from $[0,|\vec{v}_i|]$ where $\vec{v}_i$ is the deterministic portion of the velocity, \emph{i.e.}, the right hand side of (\ref{eq:ge2}).  Even this large amount of noise did not qualitatively change any of our results about rolling (or dissipated) swarms.

Finally, our model produces rolling swarms only when the corresponding free-space group has a social interaction potential in the catastrophic regime. As discussed in Section \ref{sec:model}, there is a widely-held belief that every species has a comfortable distance that is independent of group size and movement (or lack thereof) and depends only on environmental conditions; see the extended review in \cite{mkbs2003}. On the other hand, estimated locust swarm densities vary over three to four orders of magnitude \cite{r1989}. Some observations found such variation over different parts of a given swarm, and within a given swarm over a period of a few hours, as summarized in \cite{u1977}. These variations in density are not well-modeled by an H-stable potential. Of course, by using a catastrophic potential, one obtains a situation where the density of the group becomes infinite as $N \rightarrow \infty$ which is obviously unbiological. Nonetheless, we conjecture that within a range of $N$, a catastrophic potential is a reasonable modeling assumption. A similar conjecture has been made for swarming \emph{Myxococcus xanthus} cells \cite{dcbc2006}.

\section{Conclusions}
\label{sec:conclusions}

In this paper, we have constructed a discrete kinematic model for rolling locust swarms incorporating social interactions, gravity, wind, and the effect of an impenetrable boundary, namely the ground. We have studied the model using numerical simulations and tools from statistical mechanics, namely the notion of H-stability. Our simulations suggest that whether or not a swarm rolls in the presence of wind depends on the statistical mechanical properties of the corresponding free-space swarm. For a swarm that is H-stable in free space, gravity causes the group to land and form a crystalline lattice. Wind, in turn, smears the swarm out along the ground until all individuals are stationary. In contrast, for a swarm that is catastrophic in free space, gravity causes the group to land and form a bubble-like structure. In the presence of wind, the swarm migrates with a rolling motion similar to natural locust swarms observed by biologists. In the rolling swarm, all individuals land approximately the same number of times, and spend approximately the same amount of time on the ground during each landing. This captures an important feature of the real locust swarms, namely that each individual has adequate time to rest and feed.

Our study focused on two sets of parameters, one from each statistical mechanical regime. We have also conducted simulations with a limited set of other parameters and the same qualitative result was seen: swarms with a bubble structure roll, and those with a crystalline structure do not. However, this is not a proof. As mentioned in Section \ref{sec:gravity}, to further understand when a bubble forms, we are investigating a continuum analogue of our model without wind; results will be published elsewhere. The development of continuum models will also be useful because it will enable the analytical and computational study of larger swarms, as opposed to the relatively small groups we have considered here in the framework of a discrete model.

We have presented a biologically-motivated model that reproduces many features of locusts swarms observed in nature. We conclude by pointing out that the parameters in this model are chosen heuristically.  In particular, the functional form of the social interactions, while plausible, is at this point speculative. Our hope is that this work will inspire biologists to work toward gathering data that can inform and improve this model, so that eventually more quantitative comparisons with nature will be possible.

\section*{Aknowledgements}
CMT was supported by the U.S. National Science Foundation (NSF) through grant DMS-0639749 and VIGRE grant DMS-9983726. AJB was supported by NSF RTG grant DMS-0601395. SL was supported by the Harvey Mudd College Beckman Research Fund. WT was supported by NSF grant DMS-0535521. This work also received support through the Army Research Office grant W911NF-05-1-0112. We are grateful to Maria D'Orsogna, Andrea Bertozzi,  and Chris Anderson for helpful discussions.

\bibliographystyle{epj}
\bibliography{tblt2007}

\end{document}